\documentclass{aa}  

\usepackage{graphicx}
\usepackage{txfonts}
\usepackage{url} 
\usepackage{multirow}

\begin{document}

   \title{The two tails of PSR\, J2055+2539 as seen by {\slshape
       Chandra}: Analysis of the nebular morphology and pulsar proper
     motion}
\titlerunning{Two tails of PSR\,J2055+2539}


   \author{M. Marelli
          \inst{1,2}
          \and
          A. Tiengo
          \inst{1,2,3}
          \and
          A. De Luca
          \inst{1,3}
          \and
          R. P. Mignani
          \inst{1,4}
          \and
          D. Salvetti
          \inst{1}
          \and
          P. M. Saz Parkinson
          \inst{5,6}
          \and
          G. Lisini
          \inst{2}
          }
\authorrunning{M. Marelli et al.}

   \institute{INAF-Istituto di Astrofisica Spaziale e Fisica Cosmica Milano, via A. Corti 12, I-20133 Milano, Italy
             \and
             Scuola Universitaria Superiore IUSS Pavia, piazza della Vittoria 15, I-27100 Pavia, Italy
             \and
             INFN, Sezione di Pavia, via A. Bassi 6, I-2700 Pavia, Italy             
             \and
             Janusz Gil Institute of Astronomy, University of Zielona G\'ora, ul Szafrana 2, 65-265, Zielona G\'ora, Poland
             \and
             Department of Physics and Laboratory for Space Research, The University of Hong Kong, Pokfulam Road, Hong Kong
             \and
             Santa Cruz Institute for Particle Physics, University of California, Santa Cruz, CA 95064
   }


   \abstract{We analyzed two {\it Chandra} observations of PSR\,
     J2055+2539 for a total integration time of $\sim$130 ks to measure the proper motion and study the two elongated nebular features of this source. We did not detect the proper motion, setting an upper limit of
     240 mas yr$^{-1}$ (3$\sigma$ level), which translates into an upper limit on the transverse velocity of $\sim$700 km s$^{-1}$, for an assumed distance of 600 pc.
     A deep H$\alpha$ observation did not reveal the bow shock
     associated with a classical pulsar wind nebula, thus  precluding an indirect measurement of the proper motion direction.
     We determined the main axes of the two nebulae, which are
     separated by an angle of 160\fdg8$\pm$0\fdg7, using a new
     approach based on the rolling Hough transformation (RHT).
     We analyzed the shape of the first 8\arcmin\ (out of the 12\arcmin\ seen by {\it XMM-Newton}) of the brighter, extremely collimated nebula.
     Based on a combination of our results from a standard
     analysis and a nebular modeling obtained from the RHT, we find that
     the brightest nebula is curved on an arcmin scale and has a thickness ranging from $\sim9\arcsec$ to $\sim31\arcsec$ and a possible
     (single or multiple) helicoidal pattern. We could not constrain
     the shape of the fainter nebula. We discuss our results in the context of other known similar features and place particular
     emphasis on the Lighthouse nebula associated with PSR\,
     J1101$-$6101. We speculate that a peculiar geometry of the
     powering pulsar may play an important role
     in the formation of such features.
            }

\maketitle

\section{Introduction}

Rotation-powered pulsars are known to produce magnetized winds responsible for a significant fraction of the energy loss of the pulsar.
Pulsar wind nebulae (PWNe) are prominent sites of such particle acceleration detected as extended sources of nonthermal high-energy and radio emission.
The outflow morphology is influenced by the interaction with the ambient medium and by the pulsar motion. If the pulsar velocity exceeds
the sound speed in the ambient medium, a bow shock is typically present and the outflow takes a cometary-like shape \citep[for a review on PWNe, see][]{gae06}.
This classical PWN model requires an associated highly energetic pulsar ($\dot{E}\ga10^{34}$ erg s$^{-1}$) and is associated with
a bright X-ray emission surrounding the pulsar \citep[see, e.g.,][]{gae04,mcg06,kar08b}.
The PWN emission covers a wide range of energy bands, from $\gamma$-rays \citep{ack11} to the optical \citep{tem17}, 
where bow shocks are more prominent in the H$\alpha$ band \citep[see, e.g.,][]{cor93,pel02}, to radio.
However, in recent years some examples of nebulae associated with $\gamma$-ray and X-ray pulsars are adding complexity to the general picture
or even challenging this model, pointing to different physical mechanisms
being responsible for the emission of these objects \citep[][]{mar13,hui07,pav14a,kli16a,kli16b,pos17}.

The recent increase in the number of $\gamma$-ray pulsars \citep[][]{2pc}\footnote{\url{https://confluence.slac.stanford.edu/x/5Jl6Bg}}
has been crucial to study the emission phenomena related to
highly energetic pulsars, such as X-ray and radio nebulae. PSR\,
J2055+2539 (J2055 hereafter) was discovered with {\it Fermi-LAT}
as 1 of the 100 brightest $\gamma$-ray sources \citep[][]{saz10,0fgl}.
J2055 is radio-quiet, among the least energetic and oldest
nonrecycled pulsars in the {\it Fermi} LAT sample and has a spin-down energy $\dot{E}$ = 5.0 $\times$ 10$^{33}$ erg s$^{-1}$ and a characteristic age $\tau_c$ = 1.2 Myr.
\citet{mar16} analyzed a deep {\it XMM-Newton} observation of this pulsar. We found the X-ray counterpart of the pulsar, emitting nonthermal, pulsed X-rays.
Taking into account considerations on the $\gamma$-ray efficiency of
the pulsar and its X-ray spectrum, we inferred a pulsar distance
ranging from 450 pc to 750 pc. More interestingly, we found two different and elongated nebular features associated with J2055 and protruding from it.
The main, brighter feature (hereafter referred to as the main nebula) is 12\arcmin\  long and locally $<$20\arcsec\ thick and is characterized by
an asymmetry with respect to its main axis that evolves with the distance from the pulsar.
The secondary feature (hereafter referred to as secondary nebula) is fainter, shorter and broader.
Both nebulae present an almost flat brightness profile along their main axis with a sudden decrease at the end.

We analyze two new {\it Chandra} observations of the J2055 system: the first, deeper observation allows for a spatial characterization of the nebulae, while the second
observation two years later allows for the measurement  of the pulsar proper motion. Because of the accurate spectral results already reported in \citet{mar16},
based on a much more sensitive {\it XMM-Newton} observation, our
current work does not focus on the spectral study of the system.\\
The analysis of the data is described in Section \ref{data}. In
  Section \ref{results} we report our investigation into the pulsar proper motion (Section \ref{propermotion}), short-scale (Section \ref{smallsc}) and long-scale (Section \ref{largesc})
analyses of the nebulae and the analysis of the nebular shape (Section \ref{shape}).
We also observed J2055 with the 10.4 m Gran Telescopio Canarias (GTC) through an H$\alpha$ filter to search for a bow shock. This analysis is presented in Section \ref{ottico}.
A general discussion of our results in terms of physical models is reported in Section \ref{discussion}.
We developed and used a new technique for the spatial analysis of elongated features, allowing for a better analysis of diffuse emission in the X-ray band than classical methods.
This technique, based on the Hough transformation, is explained in Appendix \ref{app-ht}.

\section{X-ray observations and source detection}
\label{data}

The first {\it Chandra} \citep{gar03} observation of J2055 started on 2015 September 12 at 10:24 UT and lasted 96.8 ks.
The second {\it Chandra} observation started on 2017 September 19 at 13:56 UT and lasted 29.2 ks.
The observations were performed with the ACIS-S instrument in the Very Faint exposure mode. The pulsar was placed on the back-illuminated ACIS S3 chip.
The pointing direction and roll angle were chosen within the range allowed by viewing constraints to favor the analysis of the two tails detected with {\it XMM-Newton} \citep{mar16}.
The aim point is therefore at a distance of $\sim 1\arcmin$ from the pulsar, along the main nebula.
We observe a $\sim8\farcm5$ segment out of the total main feature that is roughly 12$\farcm$-long, distributed on the back-illuminated S3 chip and front-illuminated 
S2 chip. We also observe almost the entire roughly 250\farcs -long secondary feature.
The time resolution of the observation is 3.2 s, therefore a timing analysis of the pulsar is not possible.\\
We retrieved ``level 1'' data from the {\it Chandra} X-ray Center Science Archive
and we generated ``level 2'' event files using the {\it Chandra} Interactive
Analysis of Observations software (CIAO v.4.9)\footnote{\url{http://cxc.harvard.edu/ciao/index.html}},
as suggested by {\it Chandra} threads. 

We generated X-ray images in the 0.3--10 keV, 0.3--2 keV, and 2--10 keV energy bands using the ACIS original pixel size (0\farcs492).
The sub-bands improve the signal-to-noise of sources, allowing for a better detection of sources with thermal and with nonthermal emission, respectively.
We produced exposure maps using the standard {\tt fluximage} CIAO script.
We ran a source detection on each CCD and energy band using the {\tt wavdetect} tool\footnote{\url{http://cxc.harvard.edu/ciao/threads/wavdetect/}}; the wavelet scales ranged from 1 pixel to 16 pixel, spaced by a factor of $\sqrt{2}$.
A detection threshold of $10^{-5}$ was chosen in order not to miss faint sources. 
The performed source detection revealed the pulsar at 20$^h$55$^m$48\fs96 +25$^{\circ}$39 58\farcs78 (0\farcs3+1\farcs5 1$\sigma$ statistical plus systematic errors).
Because of the shortage of optical counterparts of the X-ray sources in the analyzed field, a boresight correction was not feasible for this observation.

\begin{figure}
\centering
\includegraphics[width=9cm]{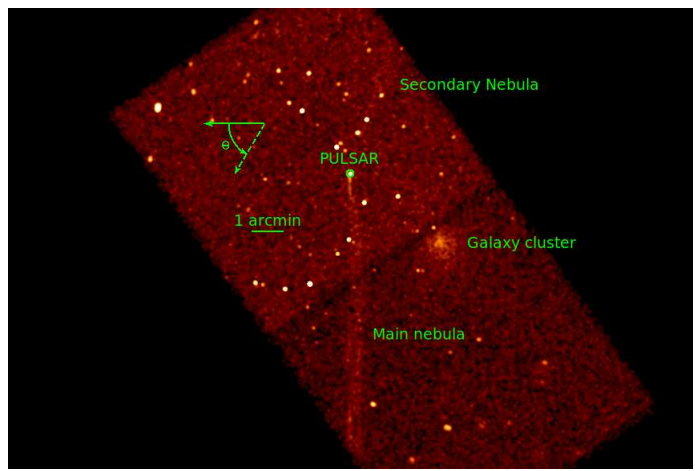}
\caption{{\it Chandra} image in the 0.3--5 keV energy band after a Gaussian smoothing. The positions of the pulsar, the nebulae, and the serendipitous
  Galaxy cluster are labeled. The image axes are  aligned in right ascension and declination with north to the top and east to the left.
  The angle $\theta$ used for the MRHT is defined such that zero is at $90^{\circ}$ toward east with respect to the north Celestial
  Meridian and is measured in the counter-clockwise direction from 0$^{\circ}$ to 180$^{\circ}$.
        }
        \label{riferimenti}
\end{figure}

The aim of our analysis is the quantitative evaluation of the statistical significance and shape of the elongated nebulae of J2055.
Based on the spectral properties of the nebulae, reported in
\citep{mar16}, we extracted events and images in the 0.3--5 keV band
to increase the signal--to--noise ratio. We used this band for all the analyses related to the nebulae.
In addition to the methods usually applied to study X-ray nebulae \citep{mar14}, owing to the low surface brightness of the
J2055 nebulae (see Figure \ref{riferimenti}) we chose to adapt and adopt a more sensitive method based on 
image pattern recognition tools, the rolling Hough transformation (RHT), which is already used in other fields of astrophysics \citep{asc10,cla14}.
The result of such an approach is the modeling of an elongated structure through linear segments. Modeling quasi-linear features using
long segments directly results in the determination of their main axes. Feature modeling through a set of short segments finds its
spatial variation and direction of substructures. For a more detailed discussion of the method and its application, see Appendix \ref{app-ht}.

\section{X-ray data analysis} \label{results}

\subsection{Pulsar proper motion} \label{propermotion}

\begin{figure}
\centering
\includegraphics[width=9cm]{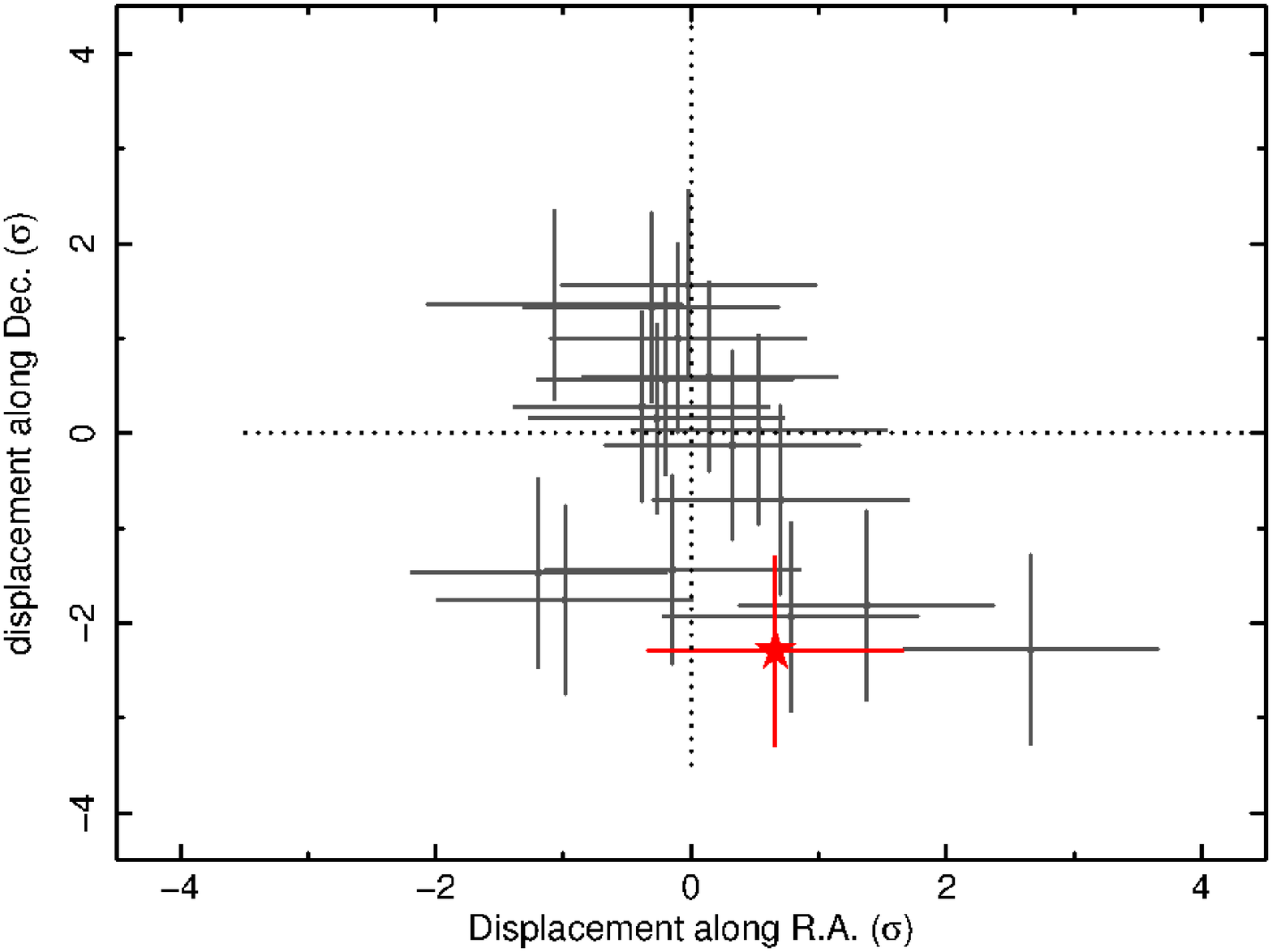}
\caption{Displacement of sources within 4\arcmin\ of the {\it Chandra}
  aim point between the two epochs of {\it Chandra} observations. The
  displacement of the pulsar is shown by the red star.
        }
        \label{fig_pm}
\end{figure}

We performed relative astrometry on our two-epoch {\it Chandra} images to search for the proper motion of the pulsar.
We used the same approach adopted to study the proper motion of PSR\, J0357+3205 \citep{del13}.
For each observation, we generated an image in the 0.3--8 keV energy range using the original ACIS pixel size (0\farcs492).
We ran a source detection using the {\tt wavdetect} task, with wavelet scales ranging from 1 to 16 pixels with a $\sqrt{2}$ step, setting
a detection threshold of 10$^{-6}$. The resulting source catalogs were cross-correlated with a correlation radius of 3\arcsec\ to extract
the list of sources detected at both epochs. The chance alignment probability of two false detections is of order $\sim10^{-5}$, based
on the density of sources in the two images.
We selected sources within 4\arcmin\ of the aim point, since point source localization accuracy
deteriorates beyond that distance because of degradation of the point spread function (PSF) with offset angle \citep[see discussion in][]{del09}. We only used
sources with a signal-to-noise larger than 4 ({\em $src_{significance}$} parameter in {\tt wavdetect} source list).
The resulting list includes 20 field sources and the pulsar counterpart. The positions of the 20 sources, which have uncertainties ranging
from 0.04 pixel to 0.3 pixel per coordinate,  were used as a reference grid for relative astrometry. We computed the best
transformation to superimpose the reference frames of the two images collected at different epochs. After rejecting 3 sources yielding large
residuals (at more than 3.5$\sigma$), a simple translation yielded a good result (see Figure \ref{fig_pm}): an r.m.s. deviation of $\sim0.3$ pixel per coordinate on the reference sources
and an uncertainty of $\sim30$ mas per coordinate on the frame registration. More complex geometric transformations are not statistically
required. We then applied the optimized transformation to the coordinates of the
pulsar counterpart to investigate its possible displacement between the two epochs.
The overall uncertainty in the pulsar displacement includes the uncertainty in the pulsar localization in each image and the uncertainty in the
image superposition. Our results are shown in Figure \ref{fig_pm}. The pulsar counterpart shows a tiny displacement, roughly pointing in the counter-direction of the secondary nebula, but its
statistical significance ($\lesssim2\sigma$) is too low to allow any claim; indeed, such a displacement is comparable to the residuals in the
positions of the reference stars after the frame registration. Thus, we assume the
corresponding yearly angular displacement, together with errors at the $3\sigma$ level, as
a conservative upper limit on the pulsar proper motion. We conclude that the upper limits on the proper motion of J2055
are $\mu_{\alpha}cos(\delta)<125$ mas yr$^{-1}$ and $\mu_{\delta}<200$ mas yr$^{-1}$.

\subsection{Small-scale analysis of the nebulae} \label{smallsc}
\begin{figure}
\centering
\includegraphics[width=9cm]{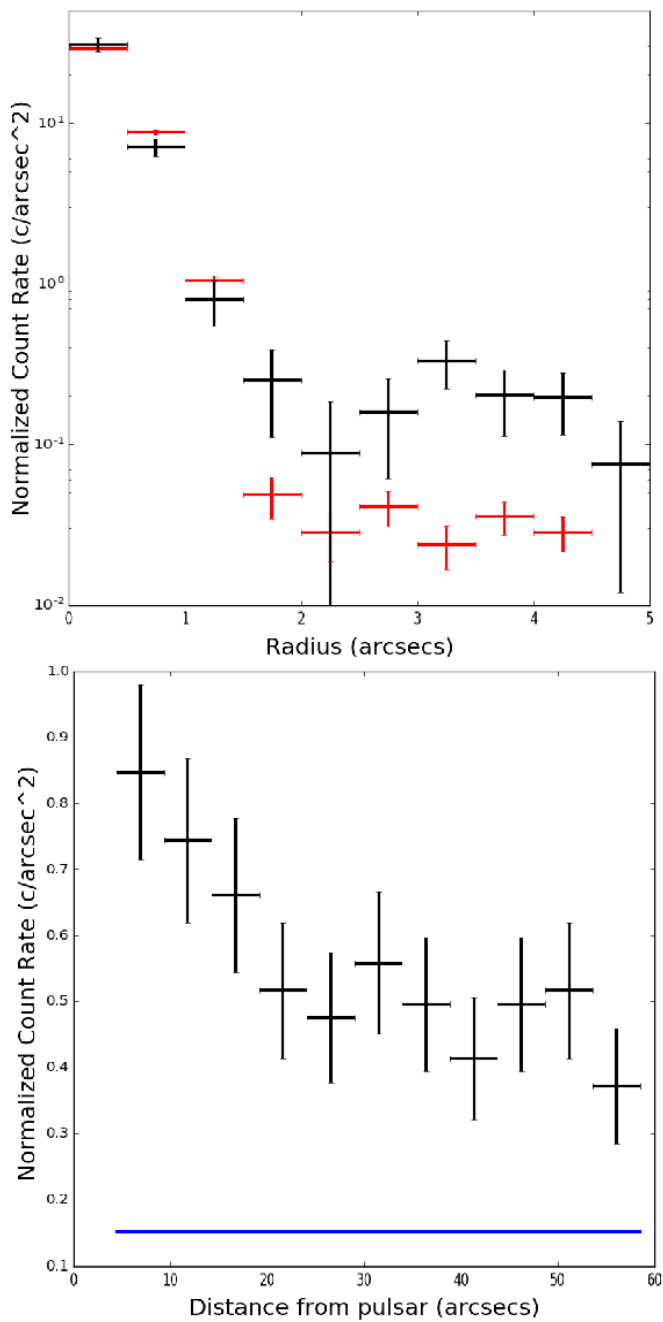}
\caption{{\it Top panel:} Circular brightness distribution of the simulated source (red) and the background-subtracted pulsar (black). Within 2'', where the contribution of
  the main nebula is negligible, the two distributions are in agreement.
  {\it Bottom panel:} Linear brightness distribution of the main
  nebula along its main axis (black) within 1\arcmin\ of the pulsar. We also show the predicted background (blue line).
  We excluded a circular region of 2\arcsec\ radius around the pulsar, where point-like emission dominates.
        }
        \label{small_scale}
\end{figure}

Using the standard CIAO tools {\tt Marx} and {\tt Chart}, we simulated a point-like source at the same position on the detector and with the same spectrum as the J2055 pulsar
but with an exposure time of 1 Ms ($\sim$ 10 times our observation). Then, we produced circular brightness distributions of the simulated source and counts actually detected from the pulsar.
The distributions for the pulsar and the simulated source are consistent, indicating no bright extended emission within 2\arcsec\ of the pulsar.
Using this method, we found that a possible bow shock located between 0\farcs5 and 1\arcsec\ from the pulsar would be detected at 3$\sigma$ with more than $\sim$75 
counts during our 96.8 ks long exposure (also considering the PSF of this source).
Taking into account the interstellar absorption along the line of sight to the source and a power-law spectrum with $\Gamma\sim2$ from \citet{mar16}, this
results in an upper limit of $\sim1.5\times10^{-14}$ erg cm$^{-2}$ s$^{-1}$ to the unabsorbed flux of a bow shock. Below 0\farcs5 the emission would not be disentangled from
the point-like emission owing to the {\it Chandra} pixel dimension (0\farcs492).\\
The previous method results in a poor description of the linear
feature protruding from the pulsar, given that the brightness is evaluated over circular annuli.
A linear brightness distribution, taken from boxes that are 10\arcmin\ thick along the main axis of the brightest nebula, detects extended emission protruding directly from the pulsar; there is a decrease of a factor $\sim$2 in brightness within 20\arcsec\
of the pulsar. Figure \ref{small_scale} shows the two distributions.

\subsection{Large-scale analysis of the nebulae} \label{largesc}

\begin{figure}
\centering
\includegraphics[width=9cm]{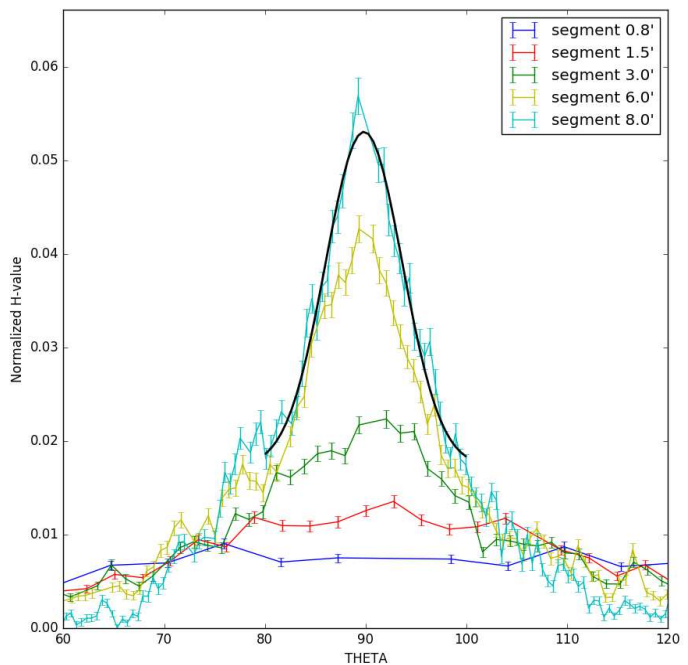}
\caption{Histogram of the angles obtained through the MRHT for a subsample of segment lengths for the main nebula analysis. Each segment is weighted using its H-value.
  The histogram integrals are normalized to one. We report the best fit of a constant plus Gaussian for the best segment using $80^{\circ}<\theta<100^{\circ}$.
        }
        \label{res_histo}
\end{figure}

\begin{figure}
\centering
\includegraphics[width=9cm]{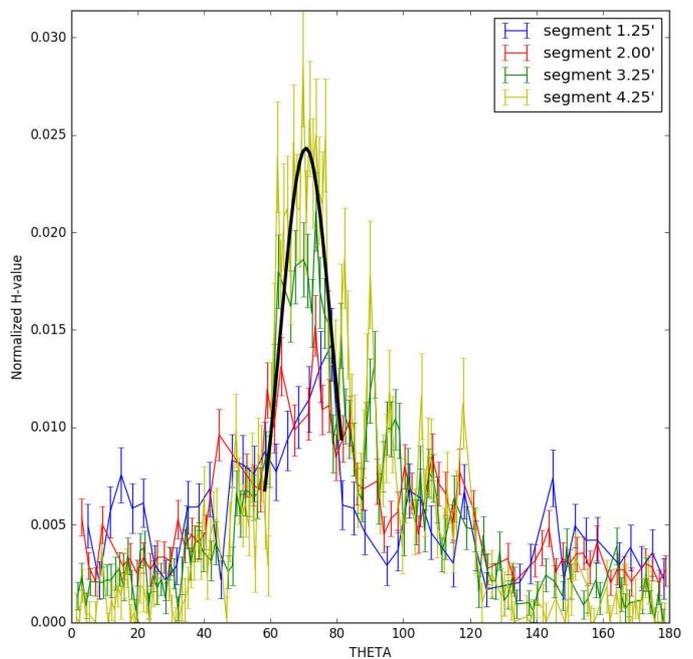}
\caption{Histogram of the angles obtained through the MRHT for a subsample of segment lengths for the secondary nebula analysis. Each segment is weighted using its H-value.
  The histogram integrals are normalized to one. We report the best constant plus Gaussian fit for the longest segment using $58^{\circ}<\theta<82^{\circ}$.
        }
        \label{res_histo_sec}
\end{figure}

In order to determine the main axes of the nebulae, we ran a modified RHT for the main nebula and secondary nebula separately (see \ref{appfin} for details).
We modeled both structures using different segment lengths $l_s$. We analyzed the distribution of the angles, ignoring the position of the segments.
For lengths comparable to the nebular length ($\sim8\arcmin$ and $\sim4\farcm2$ for the main and secondary Nebula, respectively),
the distribution of segment angles $\theta_s$ is expected to have a peak at the angle of the main axis of the nebula.
For smaller lengths, the distribution depends on the $l_s$/$t_s$ ratio, where $t_{s}$ is the mean nebular thickness
and nebular small-scale structure, as shown in Figures \ref{res_histo} and \ref{res_histo_sec}.
Using the CIAO tool {\tt mkpsfmap}, we checked that the variation of the {\it Chandra} PSF in the field of view (FoV) cannot substantially affect this result.\\
As expected, the highest value of the maximum point of the distribution of angles, $H_{\rm max}$, is reached using a segment length $l_s$ of $\sim8\arcmin$
and $\sim4\farcm2$ for the main and secondary Nebula,
respectively. Using this length, the nebula is best represented by a beam of segments with angles that peak around its main axis.
On the basis of the behavior shown, we can evaluate the angle of the main axis of the nebula by fitting the peak of the
distribution obtained for $l_s\sim$8' with a Gaussian plus a constant.
The maximum of the Gaussian distribution and its error can be taken as a measure of the angle of the main axis of the main nebula of J2055.
We obtain 89\fdg8$\pm$0\fdg1 for the main axis of the main nebula and 70\fdg6$\pm$0\fdg5 for the main axis of the secondary nebula.
The angle between the two nebulae is 160\fdg8$\pm$0\fdg7. This is in agreement within 3$\sigma$ with the result of 162\fdg8$\pm$0\fdg7 from the 
{\it XMM-Newton} observation, obtained using an independent method \citep{mar16}.
The small difference between the two results can arise from the observation of the entire main nebula (12\arcmin) in {\it XMM-Newton} data,
while we are considering only the first $8\farcm5$ visible with {\it Chandra}.
Since the line angle occurrences are not statistically independent
   because they are derived from overlapping fractions of the same image, the
  estimation of errors described above is not formally correct and the
  errors of the angle distribution histograms may be
  underestimated. We created 100 simulated data sets with a uniform,
  straight feature 15" thick, 8' long that has a brightness comparable to
  the faintest part of the real nebula (a total number of counts of
  50\% of the observed nebula); we  also considered effects introduced by
  the {\it Chandra} PSF and exposure map. After running the MRHT on these
  100 simulated images, we extracted the maximum angle, as before, and
  used these to evaluate the correct dispersion around the real
  value. The uncertainty estimates obtained through simulations were consistent
  (within 10\%) with the results obtained from the Gaussian fits applied to the real data.

\subsection{Analysis of the shape of the nebulae} \label{shape}

\begin{figure*}
\centering
\includegraphics[width=18cm]{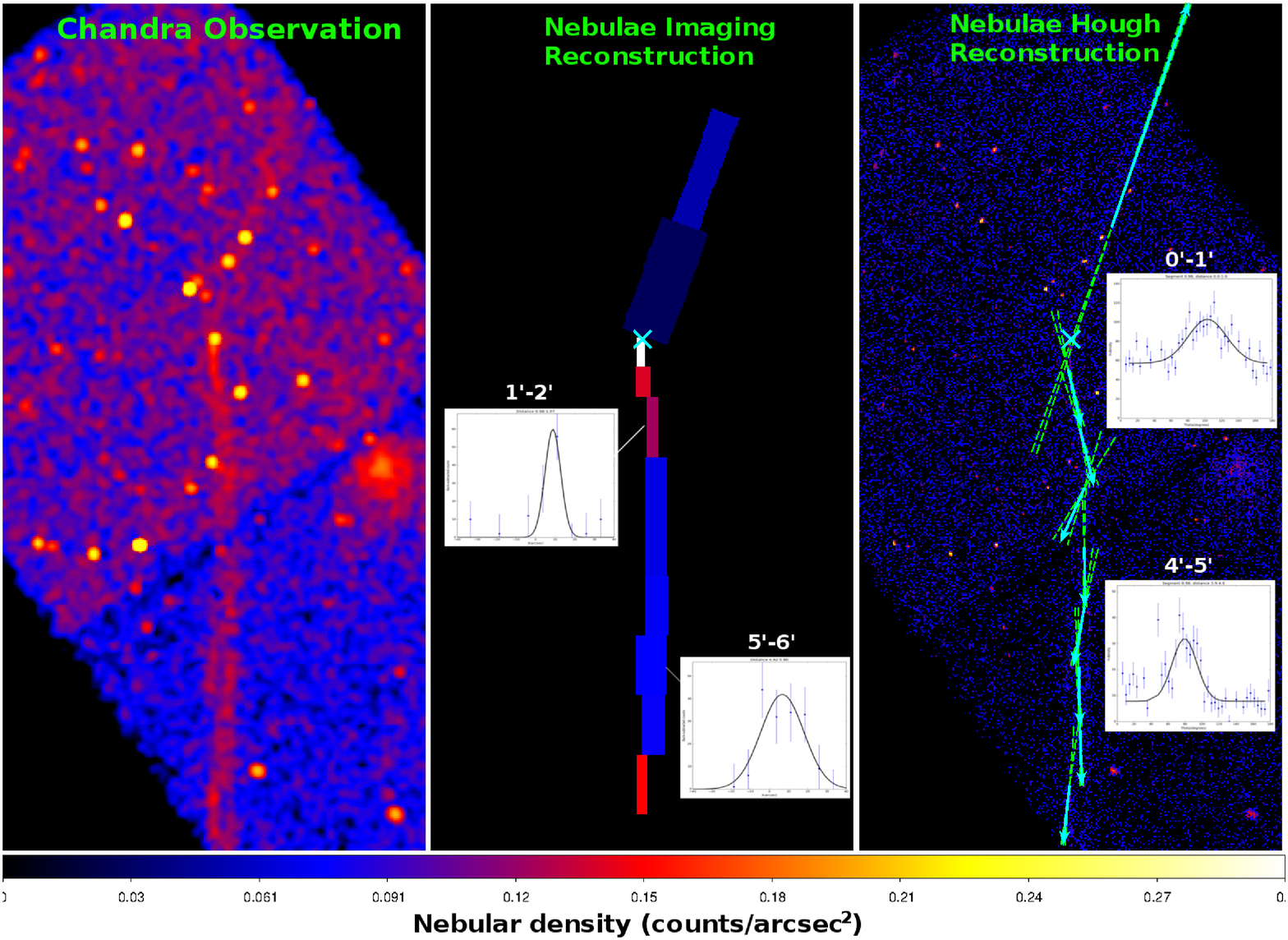}
\caption{{\it Left panel:} 0.3--5 keV {\em Chandra} image of the J2055 field. The image has been smoothed with a Gaussian filter.
  {\it Central panel:} A graphical representation of the results of the imaging analysis presented in Section \ref{shape}.
  Each box has a count distribution that varies significantly (3$\sigma$)  from the adjacent distributions. The box thickness contains 95\% of the background counts, as from Gaussian describing the distribution.
  The box intensity is the best-fitted (background-subtracted) counts density within the box. The pulsar position is indicated by a cyan X. We also report a sample of the distributions described
    in Table 1, where the distance orthogonal to the main axis of the nebula  is on the x-axis and on the background-subtracted nebular counts is on the y-axis.
  {\it Right panel:}  A graphical representation of the results of
  the MRHT analysis presented in Section \ref{results}. Each cyan vector is the best-fitted maximum of a distribution significantly (3$\sigma$) different from the adjacent distributions.
  Dotted green lines show the 1$\sigma$ error on the associated vector. The pulsar position is denoted by a cyan X. As for the central panel, we report a sample of the distributions described
    in Table 1, where the angle of significant segments from MRHT is on the y-axis.
        }
        \label{nebulas}
\end{figure*}

Having determined the main axes of the nebulae, we can quantify their variation with both classical and MRHT modeling. The first method gives
the event distribution around the axis while the second method gives the best modeling of the nebulae through linear segments.
We extracted counts from boxes centered along the main axis of each nebula, each covering
0\farcm5. We normalized for the area such that the results are in counts arcsec$^{-2}$, we applied an exposure correction, and accurately subtracted point-like sources and background.
For each box, we produced histograms reporting the background-subtracted brightness perpendicular to the main axis with bins of 7\farcs5.
For each pair of adjacent boxes, we applied a Student's t-test to the histograms to evaluate the consistency of the two distributions, and we also fitted each distribution with a Gaussian.
In the case of adjacent distributions being consistent within
3$\sigma$ (according to the t-test) and all parameters of the Gaussian
fits also being consistent within 3$\sigma$, we merged the two
contiguous boxes, obtaining a sort of adaptive binning. We iterated
this process until the distribution in each box differs significantly from the adjacent boxes.
Using such a division, we defined the thickness of the nebula in each box as the width of the rectangular region that comprises 95\% of the nebular counts in the selected box.
The effect of the {\it Chandra} PSF was taken into account by simulating PSF maps with the {\tt mkpfsmap} CIAO tool and by applying the following correction:
$\sigma_{real}=\sqrt{\sigma_{obs}^2-\sigma_{PSF}^2}$, where $\sigma_{PSF}$ is derived from a Gaussian fit of the radial profile of a simulated point-like source. Because of the small PSF size with respect to
the nebular observed thickness, the error on the Gaussian approximation of the PSF is negligible.
The results are reported in Table 1 and the central panel of Figure \ref{nebulas}.
The main nebula is thinner and brighter for the first 2\arcmin\ from
the pulsar; in this case, the nebula points toward the east.
From 2\arcmin\ to 6\arcmin\ the nebula is wider, with a fluctuating brightness, reaching the maximum
thickness between 5\arcmin\ and 6\arcmin. Between 7\arcmin\ and 8\arcmin\ the nebula is again thin almost at the level of the part nearest to the pulsar (but fainter)
and points toward west.
The secondary nebula is fainter, so that we only find a significant
difference among the first and last half, with the nearest part
thicker and/or fainter.

We followed a similar methodology using the results of the MRHT and distributions as defined in Subsection \ref{largesc}.
In this case, the Gaussian fit is performed on angular instead of count distributions and only in the central part of the peak because of its complex shape.
Again, we checked the errors on angles using simulations (100
  realizations divided into 8 groups, 1 for each
  segment). The 1$\sigma$ errors from the simulations are generally
  compatible with those from the Gaussian fit with a maximum increase of 30\%. In Table 1 and Figure \ref{nebulas} we report the results of the Gaussian fit.
The MRHT results are compatible with the features coming from the previous method (see Table 1).
For the main nebula, the angle is $\sim100^{\circ}$ for the first 2\arcmin\ from the pulsar. From 2\arcmin\  to 3\arcmin\ it is heavily affected by the presence of the CCD gap.
The main improvement from the MRHT comes from the variations seen between 3\arcmin\  and 8\arcmin. Here, the nebula significantly
fluctuates around the main axis on an arcminute scale, reaching a value of $\sim80^{\circ}$ in the last arcminute.\\
The secondary nebula is too faint to obtain any significant division. However, the MRHT approach models it with a single segment that passes through
the pulsar position (within 1$\sigma$), thus confirming the association with the pulsar and a general symmetry around the main axis.
The results of the two analyses are reported in Table 1 and the central and right panel of Figure \ref{nebulas}.

\begin{table*}
\label{table1}
\begin{center}
  \caption{Best-fit parameters of the transverse count distributions of the main and secondary nebula (see Section \ref{results} for details). Each distribution is significantly different
    ($>3\sigma$) from the previous and following distributions. The value $d_s$ indicates the position of the box with respect to the pulsar, $cd$ is the observed number of nebular counts per arcminute,
    max$_{ima}$ and $\sigma_{ima}$ are the parameters of the Gaussian fits to the count distribution while max$_{angle}$ and $\sigma_{angle}$ are the parameters of the Gaussian fits to the angle distribution. The value
    D$_{95}$ is the PSF-corrected nebular thickness containing 95\% of the background-subtracted counts. Column entries that straddle two rows indicate that they were obtained using two boxes together.
    We do not rely on angles' fit for the 2'-3' segment, due to the high contamination of the CCD gap.}
\begin{tabular}{cccccccc}
\hline
Nebula & $d_s$ & $cd$ & max$_{ima}$ & $\sigma_{ima}$ & max$_{angle}$ & $\sigma_{angle}$ & D$_{95}$\\
- & arcmin & arcmin$^{-1}$ & arcsec & arcsec & deg & deg & arcsec\\
\hline
Main & 0-0.5 & 180$\pm$18 & -3.1$\pm$1.6 & 4.7$\pm$1.2 & \multirow{2}{*}{103.0$\pm$3.0} & \multirow{2}{*}{32.4$\pm$5.3} & 9.2$\pm$2.4\\
Main & 0.5-1 & 202$\pm$18 & -1.2$\pm$1.7 & 8.1$\pm$2.5 &  &  & 16.1$\pm$5.0\\
Main & 1-2 & 86$\pm$60 & 8.9$\pm$1.1 & 5.7$\pm$1.4 & 103.7$\pm$4.4 & 36.8$\pm$8.7 & 11.3$\pm$2.8\\
Main & 2-3 & 143$\pm$49 & \multirow{2}{*}{12.3$\pm$1.8} & \multirow{2}{*}{10.8$\pm$2.4} & 66.0$\pm$7.5 & 36.8$\pm$12.7 & \multirow{2}{*}{20.7$\pm$4.8}\\
Main & 3-4 & 53$\pm$46 &   &   & 90.4$\pm$1.7 & 4.0$\pm$1.9 & \\
Main & 4-5 & 92$\pm$46 & 13.5$\pm$2.2 & 12.1$\pm$3.0 & 79.4$\pm$2.6 & 20.5$\pm$3.9 & 24.1$\pm$6.0\\
Main & 5-6 & 121$\pm$47 & 6.8$\pm$3.4 & 15.8$\pm$4.6 & 91.6$\pm$1.8 & 27.4$\pm$2.7 & 31.4$\pm$9.2\\
Main & 6-7 & 84$\pm$46 & 9.1$\pm$2.6 & 11.9$\pm$3.6 & 91.9$\pm$1.8 & 22.9$\pm$2.6 & 23.6$\pm$7.2\\
Main & 7-8 & 124$\pm$46 & -1.7$\pm$1.2 & 6.3$\pm$1.7 & 83.8$\pm$1.9 & 32.3$\pm$3.1 & 12.0$\pm$3.4\\
Secondary & 0-2 & 87$\pm$31 & -3.0$\pm$6.1 & 24.4$\pm$9.1 & \multirow{2}{*}{71.0$\pm$0.6} & \multirow{2}{*}{10.0$\pm$1.7} & 48.8$\pm$18.2\\
Secondary & 2-4 & 113$\pm$31 & -6.9$\pm$2.4 & 14.2$\pm$3.4 &  &  & 28.2$\pm$6.8\\
\hline
\end{tabular}
\end{center}
\end{table*}

\section{Optical observations and data analysis} \label{ottico}

To search for alternative evidence of the pulsar motion and determine its direction
in the plane of the sky, we looked for an arc-like emission structure from ionized hydrogen produced at the termination shock of the pulsar wind as it moves supersonically in the interstellar medium (ISM).
To detect such a bow shock, we obtained deep observations of the J2055 field with the GTC on August 22 and 23, 2015 under program GTC12-15A.
We observed J2055 with the Optical System for Imaging and low Resolution Integrated Spectroscopy (OSIRIS).
The instrument is equipped with a two-chip E2V CCD detectors with a nominal FoV of $7\arcmin \times 7\arcmin$.
The pixel size of the CCD is 0\farcs25 in the $2\times2$ binning mode. We took three sequences of 15 exposures in the H$\alpha$
filter ($\lambda=6530$\AA; $\Delta \lambda=160$\AA) with exposure time of 155 s each to minimize the saturation of bright stars in
the field and remove cosmic ray hits. We also took one sequence of 15 exposures (also of 155 s each) in the $r'$
($\lambda=6410$ \AA; $\Delta \lambda=1760$\AA) filter. In order to
cover as much of the main nebula as possible, the pointing coordinates were offset by 2\arcmin\ to the east and 2\arcmin\ to the south. Observations were performed in
gray and clear sky conditions with an average airmass of 1.22 and a seeing of $\sim 0\farcs9$.  Short (0.5--3 s) exposures of standard star fields \citep{smi02}
were also acquired for photometric calibration, together with twilight sky flat fields.

We reduced our data (bias subtraction, flat-field correction) using
standard tools in the {\sc IRAF}\footnote{IRAF is distributed by the National Optical Astronomy Observatories, which are operated by the Association of Universities
  for Research in Astronomy, Inc., under cooperative agreement with the National Science Foundation.} package {\sc ccdred}.
Per each filter, single dithered exposures were then aligned, average-stacked, and filtered for cosmic rays using the {\sc IRAF} task {\tt drizzle}.
We computed the astrometry calibration on the GTC images using the {\em wcstools}\footnote{{\texttt http://tdc-www.harvard.edu/wcstools}},
matching the sky coordinates of stars selected from the Two Micron All Sky Survey (2MASS) All-Sky Catalog of Point Sources \citep{skr06}
with their pixel coordinates computed by {\em Sextractor} \citep{ber96}.  After iterating the matching process applying a $\sigma$-clipping
selection to filter out obvious mismatches, high-proper motion stars,
and false detections, we ended up with an overall accuracy of $\sim$0\farcs2 on our absolute astrometry.

We did not detect any arc-like structure that could be identified as a bow shock associated with the pulsar in the stacked H$\alpha$ image.
We computed a $3\sigma$ surface brightness limit on the undetected bow shock from the rms of the sky background sampled within a radius
of 5\arcsec\ of the pulsar position, conservatively set to account for the uncertainty on the stand-off distance between the pulsar and
the bow-shock apex (see Section \ref{discussion}). We note that the pulsar falls $\sim 2\arcsec$ southwest from a star \citep[$r'=20.34\pm0.04$; ][]{mig18},
which increases the rms of the sky background near to it. However, this would not dramatically affect our limit unless the bow-shock axis and the pulsar
proper motion direction were pointing at the star at a position angle of $\sim 55^{\circ}$ (incompatible with that of the main and secondary nebulae),
its stand-off distance were of the same order of the pulsar/star separation, and its angular size were smaller  than the image PSF. We converted instrumental
magnitudes to physical flux units by cross-calibrating the H$\alpha$ image against the $r'$ band image. To this aim, we matched about 500 stars in common
between the two images (1\arcsec\ matching radius) detected by {\em Sextractor}  at $5\sigma$ above the background, after filtering saturated stars, blends,
and stars too close to the CCD edges. We compared the instrumental magnitudes of the stars through aperture photometry using an aperture
of 2\farcs5 diameter, i.e., about three times the average seeing.  We applied a linear fit between the instrumental magnitudes to obtain the relative photometry
transformation after applying a $\sigma$ clipping to filter out mismatches and outliers. This turned out to be accurate to within $\pm 0.2$ magnitudes.
By applying the $r'$ band night zero points and the airmass correction using the atmospheric extinction coefficients for the La Palma
Observatory\footnote{{\texttt  www.ing.iac.es/Astronomy/observing/manuals/ps/tech\_notes/tn031.pdf}} we then obtained a  $3\sigma$ surface brightness
limit of $\sim 2.3 \times 10^{-17}$ erg cm$^{-2}$ s$^{-1}$ arcsec$^{-2}$ in the H$\alpha$ filter, which we assume as the upper limit on the surface brightness
of the bow shock within a 5\arcsec\ distance from the pulsar.

\section{Discussion}
\label{discussion}

We analyzed two new {\it Chandra} data sets of the J2055 system, focusing our analysis on the small- and large-scale shape of the nebular features
protruding from the pulsar and on the search for the pulsar proper motion. We based the imaging analysis on the combination of a classical method
and a newly developed method for the analysis of linear features called the
modified rolling Hough transformation (MRHT).

We do not find evidence of proper motion down to 240 mas yr$^{-1}$ (3$\sigma$ level), which translates into an upper limit on the transverse velocity of $\sim$700 km s$^{-1}$ at 600 pc.
If we consider the system to be 600 pc distant, the angular lengths of 12\arcmin\ for the main nebula and 250\arcsec\ for the secondary nebula \citep{mar16}
translate into projected physical lengths of 2.1 and 0.7 pc, respectively.
Both values are consistent with other observed X-ray synchrotron nebulae, which show projected lengths up to a few parsecs \citep[see, e.g.,][]{kar08a,kar17}.
Considering our upper limit on the transverse velocity, the pulsar covered the lengths of the nebulae in $t_p>$3 kyr or $t_p>$1 kyr, respectively.
Taking into account the classical sychrotron-emitting nebula model, the electrons are accelerated at the termination shock and injected into the
cavity produced by the fast-moving pulsar.
As discussed in \citet{mar16}, using optimistic values of ambient magnetic field and Lorentz factor for electron acceleration we obtain a synchrotron cooling time of the
emitting electrons of $\tau_{sync}$ $\sim$ 100 yr $<<t_p$. In order to reach the physical extension of the nebula, we thus need a bulk flow speed of the emitting particles
$>$20000 km s$^{-1}$ and $>$7000 km s$^{-1}$, respectively, for the two nebulae. The first value is only marginally consistent with results in the literature,
while the second is fully consistent \citep{kar08a}. Also, the lack of spectral variation in the main nebula \citep{mar16} points to a much
higher bulk flow velocity or a reacceleration mechanism along this nebula. Thus, the classical synchrotron nebula model is disfavored in explaining
the main nebula emission.\\
For classical synchrotron nebulae, we expect a relatively bright diffuse emission surrounding the pulsar, where
the emission from the wind termination shock is brightest, as observed in all the other known cases \citep[e.g.,][]{kar08b}.
Although there is not a clear correlation, typical luminosities of the ``bullet'' component versus the ``tail'' component are $L_{\rm bull}/L_{\rm tail}>$0.1.
Assuming the standard relations from \citet{gae06}, the distance between the pulsar
and the head of the termination shock is expected to be $r_s = (\dot{E}/4\pi\rho_{ISM}v_{PSR}^2)^{1/2}$,
where $\rho_{ISM}$ is the ambient density and $v_{PSR}$ is the pulsar velocity. Taking into account a typical ambient density (0.1 atoms cm$^{-3}$) and our upper-limit velocity at 600 pc,
this would place the shock at $\sim0\farcs5$ or further in case of a lower pulsar velocity.
On a sub-arcsec angular scale, we do not find any evidence of a termination shock at $\gtrsim$0\farcs5 with an unabsorbed flux $>1.5\times10^{-14}$ erg cm$^{-2}$ s$^{-1}$.
This makes the bullet component of the nebula fainter than expected if the main nebula is a typical synchrotron nebula, with $L_{\rm bull}/L_{\rm tail}<$0.05.
As reported in \citet{mar16} the low energetics of the powering pulsar ($\dot{E}$ = 5.0 $\times$ 10$^{33}$ erg s$^{-1}$) also disfavor
the model of a classical synchrotron-emitting nebula for the bright main feature.
The luminosity of the secondary, fainter nebula instead is compatible with a classical PWN, both
regarding the luminosity ratio $L_{\rm bull}/L_{\rm tail}<$0.25 as well as the comparison of the nebula luminosity with the pulsar energetics.

The deep upper limits on the surface brightness of any undetected diffuse structure in our $H_{\alpha}$ images, coupled to the upper limit on the
proper motion of the pulsar, allow us to constrain the properties of the ISM in the region surrounding the pulsar. Assuming
the distance to the pulsar to be smaller than 900 pc, and the the pulsar space velocity vector to point within $25^{\circ}$ of the plane of the sky, the
scaling relations proposed by \citet{cor93} and \citet{cha02} \citep[see also ][]{pel02} imply a neutral hydrogen
fraction lower than 0.2, suggesting that the pulsar should be moving in the warm or hot component of the ISM. Some contribution to the ionization of
the ISM could of course be due to the UV flux from the pulsar itself.

Through the MRHT approach we determined the main axes of the two features, separated by 160\fdg8$\pm$0\fdg7. The evaluation of the main axes
of the two nebular structures also allowed us to produce results from classical methods.
The main nebula is tight and has a 95\% thickness $9\arcsec<t_{95}<31\arcsec$ because this feature is thinner nearby and far from the pulsar;
this feature is evaluated on an arcminute-scale along the first 8\farcm5 of the observable nebula (out of the 12\arcmin\ seen by {\it XMM-Newton})
and also directly protrudes from the pulsar on an arcsecond scale.
The brightness profile changes with distance on an arcmin scale because it is wider for distance $2\arcmin<d<7\arcmin$ and brighter near the pulsar ($d<1\arcmin$) and far from it ($7\arcmin<d<8\arcmin$).
On a similar scale, the direction of the nebula also changes in the range $71^{\circ}<\theta<92^{\circ}$ for $3\arcmin<d<8\arcmin$,
where the angle $\theta$ is defined such that the zero is at $90^{\circ}$ due east with respect to the north Celestial Meridian
and is measured in the counter-clockwise direction; this nebula follows a snake-like shape.

Normal to the main axis, the distributions of the nebular counts are
not uniform with distance from the pulsar. Moreover, both the
classical analysis and the MRHT detect hints of a split in the structure of the wider segments, $5\arcmin<d_{\rm pulsar}<6\arcmin$ and $6\arcmin<d_{\rm pulsar}<7\arcmin$.
For the classical method, this comes from the wider peak in the distribution between $5\arcmin<d_{\rm pulsar}<6\arcmin$ (see Table 1 and panel in Figure 6),
which is not supported by a similar widening in the angles and thus points to parallel structures.
For the MRHT method, this is also apparent in Figure A.3, left panel, when we use a segment length of 1'.
This resembles a single (or multiple, if the split is confirmed) helicoidal pattern, although we cannot exclude an irregular short-scale-bumped shape.
Given the fainter and less collimated nature of the secondary
nebula, compared to the main nebular, apart from its direction, we can only infer a hint of narrowing with distance and a roughly symmetric profile
perpendicularly to its main axis. Only future deeper {\it Chandra} observations will be able to better define the nebular structures.\\
We found four other pulsars in the literature associated with elongated parsec-long features misaligned with the proper motion direction.
The pulsar PSR\, B2224+65 \citep{hui07}
is associated with a long, extended X-ray feature whose orientation deviates by $\sim$ 118$^{\circ}$ from a classical bow-shock nebula (the Guitar Nebula), seen in H$\alpha$
in the counter-direction of the proper motion. An X-ray counter-feature of the X-ray nebula was also detected \citep{joh10}, albeit substantially fainter and shorter than the main nebular.
PSR\, J1101$-$6101 \citep{hal14} is associated with the long, collimated Lighthouse Nebula,which deviates by $\sim104^{\circ}$ from a classical X-ray
PWN in the counter-direction of the presumed proper motion direction \citep{pav14a}. This nebula is well modeled by a (multi)helical pattern.
A short, faint counter-feature of the Lighthouse Nebula is also detected in X-rays. This system also comprises a classical PWN, counter-aligned with the
pulsar proper motion direction and less collimated than the other two structures.
PSR\, J1509$-$5850 is associated with a parsec-long X-ray PWN \citep{kar08a} extending in the counter-direction of the proper motion. Recently, a fainter elongated, parsec-long feature
has been detected in X-rays, deviating by $\sim147^{\circ}$ from the brighter nebula \citep{kli16a}.
Finally, PSR\, B0355+54 is associated with the Mushroom X-ray PWN, oriented in the counter-direction of the proper motion. Hints of two faint, short,
counter-aligned features have been observed in X-rays, one of which is twice the length of the other, almost orthogonal to the pulsar proper motion direction \citep{kli16b}.\\
Taking into account the rotational energy loss that is converted into
the luminosities of the misaligned features, we obtain values 0.0004$<L_{\rm bol}/\dot{E}<$0.07
(with a value of 0.0064 for J2055) and typical projected lengths 0.7$<l_{proj}<$7.7 pc (with a value of $\sim$2.1 pc at 600 pc for J2055),
while the energetics of the powering pulsars span three orders of magnitude (from $\sim10^{33}$ to $\sim10^{36}$ erg s$^{-1}$). For a general view of such nebulae
see Figure 9 in \citet{kar17}.\\
In the case of J2055, the possible helical morphology of the main nebula resembles the Lighthouse Nebula, among the four cited cases.
The secondary nebula can hardly be interpreted as a counter-feature of
the main nebular, given the misalignment of the two features, in particular taking into account only the first part of the main nebula.
Also, all the other known counter-features are collimated, and the secondary nebula is not.
It is more consistent, given the energetics, direction, and low collimation, with a classical PWN as in the case of the classical PWN of the Lighthouse system.

To explain the nature of the motion-misaligned, collimated features such as the main nebula of J2055, two models are commonly invoked.
\citet{ban08} considered a scenario in which high-energy particles leak from the termination shock apex into the ISM and
travel along the ordered ISM magnetic field. The main problem with this scenario is the lack (or faintness) of counter-features,
which means that there is a preferred direction for the escape of particles. The cause of such an asymmetry is unclear.
Although a Doppler effect could be invoked, the discussion is still open \citep[see, e.g.,][]{kli16a,kar17}.
As reported in \citet{pav14a}, the helical pattern of the J2055 nebula weakens an interpretation based on this model.
The second model invokes a powerful, collimated ballistic jet from the pulsar poles. Such jets should be bent by the ram pressure of the oncoming
ISM, while the lack (or faintness) of the counter-feature is explained in terms of Doppler boosting. Such an interpretation is particularly intriguing
for the Lighthouse Nebula, whose shape is well modeled by a (single or
multiple) helical pattern. This is ascribed to the precession of the pulsar (possibly due to its oblateness)
or kink instabilities in the jet. A comprehensive discussion of this
model is given in \citet{pav14a,pav16}. The shape of the J2055 main
nebula is reminiscent of the Lighthouse Nebula, only coming from a
pulsar that is $\sim200$ times less energetic and $\sim$10 times nearer.\\
For both nebular models, a key role could be played by the geometry and magnetic field configuration of the powering pulsar.
Unlike PSR\, J1101$-$6101, J2055 is a strong emitter of $\gamma$-ray radiation and models have been developed to describe its magnetic field configuration.
The second {\it Fermi} pulsar catalog \citep{2pc} reports that J2055 is one of the very few $\gamma$-ray pulsars with a strong detection of off-pulse emission characterized by a spectral
cutoff: this emission probably originates from its magnetosphere, opening a new conundrum for magnetospheric emission models.
Such weakly pulsed emission should be rare as it is expected only for nearly aligned pulsators seen at high inclinations using classic outer-gap models \citep{rom10}.
\citet{pie15} used different $\gamma$-ray and radio emission geometries to fit $\gamma$-ray and radio light curves of {\it Fermi} pulsars, where
outer-gap and one-pole caustic emission models are favored in explaining the pulsar emission pattern of {\it Fermi} pulsars.
Among these, J2055 is best fitted using models where $\gamma$-ray emission comes from near the pulsar surface, such as the Polar Cap \citep{mus03} and Slot Gap \citep{mus04} models,
while outer-magnetosphere models fail to predict the light curve shape correctly. We note that such models are usually disfavored in
describing pulsar emission \citep[see, e.g.,][]{2pc,pie15}; there are some
notable exceptions \citep[see, e.g., PSR\, J0659+1414][]{wel10}  and J2055 could well be one of these few exceptions.\\
This suggests that J2055 could have a rare geometry, although the exact nature of that shap is still unclear. This geometry could be the key to understanding why only a small
number of pulsars show such powerful, collimated features, and how they are produced.

More observations, as well as the application of new analysis techniques and models, could help us to understand the shape and spectrum of such misaligned nebulae.
Deep investigations of pulsars with geometrical configurations similar to J2055 would also greatly help to develop models for such features, clarifying the
role of the magnetic field of the pulsar in the creation of such powerful, extended sources.

\section{Conclusions}

Using two recently obtained {\it Chandra} observations of PSR
J2055+2539, we set an upper limit on its proper motion of 240 mas
yr$^{-1}$, which translates into an upper limit on its transverse
velocity of $\sim$700 km s$^{-1}$ at 600 pc.  We found no evidence
of bow shocks, either in the X-rays or in H$\alpha$, at scales
$\gtrsim$0\farcs5. Two almost-linear features protrude from the
pulsar.  The main, brighter nebula is highly collimated and its
shape is reminiscent of a (multi)helicoidal pattern, resembling the
long, motion-misaligned feature seen in the Lighthouse Nebula. Because of its brightness, shape, lack of a bow shock, and low pulsar
energetics we rule out the main nebula as a classical PWN.  Four
other known systems present long, very collimated nebulae misaligned
with the pulsar proper motion, one of which has a (multi)helicoidal
shape. We conclude that the main nebula is produced by the same
physical process as these four. We speculate that this process might
be related to a peculiar geometry of the magnetosphere of the
powering pulsar, which is confirmed in the case of J2055  via its
$\gamma$-ray properties.\\
The secondary nebula is consistent with a classical PWN model, based
on the pulsar energetics and nebular luminosity, and on the
lack of a detected bow shock. However, only future
observations including the detection of the pulsar proper motion
can confirm our hypothesis.

\begin{acknowledgements}

We thank Nanda Rea for the use of the data of GTC under program GTC12-15A and for the useful discussion. We thank the referee for the useful comments that really improved the paper in many ways.
This work was supported by the Fermi contract ASI-INAF I-005-12-0.
This work was supported by the IUSS contract for the project $''$Studio del cielo ad alte energie: variabilit\'a del cielo nei raggi X oltre EXTraS e nei raggi gamma in vista del CTA$"$.
RPM acknowledges financial support from an INAF "Occhialini Fellowship".
Support for this work was provided by the National Aeronautics and Space Administration through Chandra Award Number GO5-16076X issued by the Chandra X-ray Center, which is operated by the Smithsonian Astrophysical Observatory for and on behalf of the National Aeronautics Space Administration under contract NAS8-03060.

\end{acknowledgements}

\begin{appendix}

\section{Modified rolling Hough transformation with application to {\it Chandra} data} 
\label{app-ht}
  
The Hough transformation (HT) was first used for the detection of complex patterns in bubble
chamber photographs \citep{hou62}. Soon, it proved to be a powerful tool for the detection of linear patterns and has found many
applications in image processing \citep{ill88}.
\citet{cla14} adapted the HT in a rolling version (RHT) to
study linear HI features in the diffuse Galactic ISM. This method
was recently used also by \citet{jel08}.
We adapted the RHT for a global characterization of X-ray observations, which usually have lower statistics than optical observations.
Through the comparison of the input image with an expected, simulated background image,
we are able to extract the significance of each segment in the pattern.
In order to apply our MRHT algorithm, we also need to flatten the {\it Chandra} background
through the careful handling of contaminating celestial sources, exposure-map variation, and different chip illumination.
The MRHT approach and the flattening of {\it Chandra} background applied to our observation are described in Appendix \ref{app-ht-flat}.
As a final step, the MRHT produces a list of segments with a significance $S_s>3.5\sigma$ characterized by the following six parameters:\\
- $l_s$, the length of the segment (that is given as input)\\
- $(x_s, y_s)$, the position of the center of the segment\\
- $\theta_s$, the angle of the segment, from 0$^{\circ}$ to 180$^{\circ}$, 
defined such that the zero is at $90^{\circ}$ due east with respect to the north Celestial Meridian and measured in the counter-clockwise direction  (see Figure \ref{riferimenti})\\
- $S_s$, the significance of the segment in units of standard deviations\\
- $H_s$, the density of exposure-corrected events falling on the segment in counts/pixel units

\subsection{ Hough transformation}

Our implementation of the Hough transformation follows that of \citet{hou62}.
In this basic implementation, given a Cartesian coordinate space (x-y), a straight line in this space can be described through the angle $\theta$ of its normal passing through
the origin and the minimum Euclidean distance $\rho$ from the origin,  as follows:

\begin{equation}
\rho = x cos \theta + y sin \theta
\end{equation}

Each single ($x_i$, $y_i$) event pixel in the image (x-y) space can be, therefore, transformed into a sinusoidal curve in the $\rho-\theta$ parameter space,
where  $\rho = x_i cos \theta + y_i sin \theta$.
The sinusoid beams that result from transformed colinear points in the image space have a common
point of intersection ($\rho_0,\theta_0$) in the parameter space. The parameters ($\rho_0,\theta_0$) describe the line that passes through such colinear points in the image space. 
Usually the $\rho$-$\theta$ space is quantized to reduce the computational burden considerably.
The Hough transformation thus stores in a ($\rho$, $\theta$) array the number of events in the image space that contribute to each pixel in the parameter space
H($\rho,\theta$), basically, the number of events for each line in the image space.
Thus, the problem of detecting colinear points in the image space can be converted to the problem of finding bright pixels in the parameter space.
Setting an intensity threshold H$_{thresh}$ in the parameter space, the most prominent lines in the image space can be extracted and represented in the image space.
The obvious limitations of this method are\\
- it can only be used to detect straight, bright lines that significantly contribute to the entire image space;\\
- it cannot assign to each line the probability of being spurious, only allowing for a user-selected intensity cut; and\\
- it is not sensitive to background spatial variations.\\

\subsection{Rolling Hough transformation}

The rolling Hough transformation performs a Hough transformation N times on circular sub-images, extracted from a circular region with a given diameter D and center ($x_c, y_c$).
For each sub-image the parameter space is limited to $\rho$ = 0, so that the $\rho-\theta$ space is reduced to a one-dimensional space on $\theta$, for each sub-image defined by ($x_c, y_c$).
Only pixels in each $(\theta)$ sub-space (extracted from D-length segments centered in ($x_c,y_c$) in the image space) with intensity H($\theta,x_c,y_c$)
over a user-set intensity threshold H$_{thresh}$ are considered.
A visualization of the linear structures identified by the RHT, the back-projection H(x,y) is obtained by integrating H($\theta,x_c,y_c$) over $\theta$.
The RHT approach has two main limits, when applied to the X-ray observations:
First, to extract lines, the user must choose a given intensity, or intensity percentage that is not based on a statistical analysis. Second, this method requires a good statistics (e.g., in a Gaussian approach) for each circle analyzed with the Hough transformation, thus requiring sufficiently large circular regions in the sky.

\subsection{Modified rolling Hough transformation}

Following the same sub-selection of D-diameter circular images
approach as in the RHT, we evaluate the threshold based on the
probability P$_H$ that the line in the image space is spurious, rather than on a given intensity threshold H$_{thresh}$.
This is possible thanks to Monte Carlo simulations run on a ``flat background'' image. We apply the RHT transformation on the simulated images to extract
the mean histogram $f(H)$ of H-values in the case of a source-free background. We can thus link the H-value($\theta,x_c, y_c$), basically the number of events on the segment in the image space
defined by ($\theta,x_c, y_c$),  with the probability of being spurious, defined as\\

\begin{eqnarray}
  P(H) = \frac{int_{H}^{\infty} f(H) dH}{int_{0}^{\infty} f(H) dH}
.\end{eqnarray}

It is straightforward to convert $P(H)$ into a significance $S(H)$ (expressed in $\sigma$).
Lines in the image space below a probability threshold P$_H$ are discarded, while the others are stored and contribute to the back projection.
This method is applied several times for a grid of extraction diameters D, thus allowing for the detection and characterization of segments of different lengths.
A normalization that takes into account the segment length is applied (using $H/D$ instead of $H$).\\
We note that this method requires a background that is as flat as possible. The probability is evaluated globally, thus a background spatial
variation would result in a spatially variable sensitivity to features and, therefore, spurious detection of background features.
Part of our work involved correcting instrumental features of the {\it Chandra} images, such as exposure map variations and different characteristics of individual CCDs,
and the removal of contaminants such as point-like sources.

\subsection{Flattening the {\it Chandra} background} \label{flatten}
\label{app-ht-flat}

In order to apply our modified RHT algorithm, we need to decrease the spatial variations of the {\it Chandra} background in our observations.
There are four main sources of contamination as follows:\\
- Celestial sources: Both the point-like sources and the detected Galaxy cluster \citep{mar16} could affect the characterization of the nebulae.\\
- Exposure map: Variations in the exposure map result in a variation of the background count rate, affecting the global H-values distribution
and resulting in a decrease of the HT sensitivity and detection of spurious, background-related lines.\\
- Chip illumination: We perform our analysis on S2 and S3 {\it Chandra} CCDs, front-illuminated and back-illuminated, respectively; this results in a different
background count rate level, and thus leads to the same problems coming from exposure map variations.\\
- Unexposed areas: The presence of unexposed areas heavily affects the global H-values distribution and could hamper the analysis using long segments.\\

For the HT we worked on the 0.3--5 keV images and exposure maps.
We cheesed them, excluding elliptical regions around each source containing 99\% of expected source counts, as extracted by {\tt wavdetect}.
We also excluded the galaxy cluster in the S2 CCD using an ad hoc circular region, evaluated using both {\em Chandra} and {\em XMM-Newton} data.

We converted the standard exposure maps into fractional exposure maps, with each pixel (x,y) value $E_{frac}$ defined as

\begin{eqnarray}
  E_{frac}(x,y) = \frac{E(x,y)}{max(E)}
.\end{eqnarray}

Vignetting effects within single chips are taken into account using vignetted exposures.
Pixels with very low ($<$0.3) fractional exposure (3.1 and 2.6 arcmin$^2$ on a total of 74.6 and 74.6 arcmin$^2$ for CCD S2 and S3, respectively)
were set to zero: for very low statistics the exposure correction diverges.
Therefore these regions are excluded both from fractional exposure maps and images.
To correct for exposure map variations, image pixel values are multiplied by the corresponding pixels values in the percentage exposure map.

We selected ad hoc source-free regions with high ($>$0.9), uniform exposure (22.3 and 50.9 arcmin$^2$ for CCD S2 and S3, respectively).
We extracted the (exposure-corrected) background count distribution from these regions.
This distribution, after an area correction, was used to evaluate the probability of adding a new count in each pixel of cheesed area (where the exposure map of the CCD is not null).

Then, we co-added the two CCDs image maps, correcting for the different chip illumination. The two background distributions extracted from the two chips are compared using a Gaussian fit.
We randomly added counts to the pixels of CCD 2 (front-illuminated, with the lower background) to correct for the different distributions.

\begin{figure*}
\centering
\includegraphics[width=18cm]{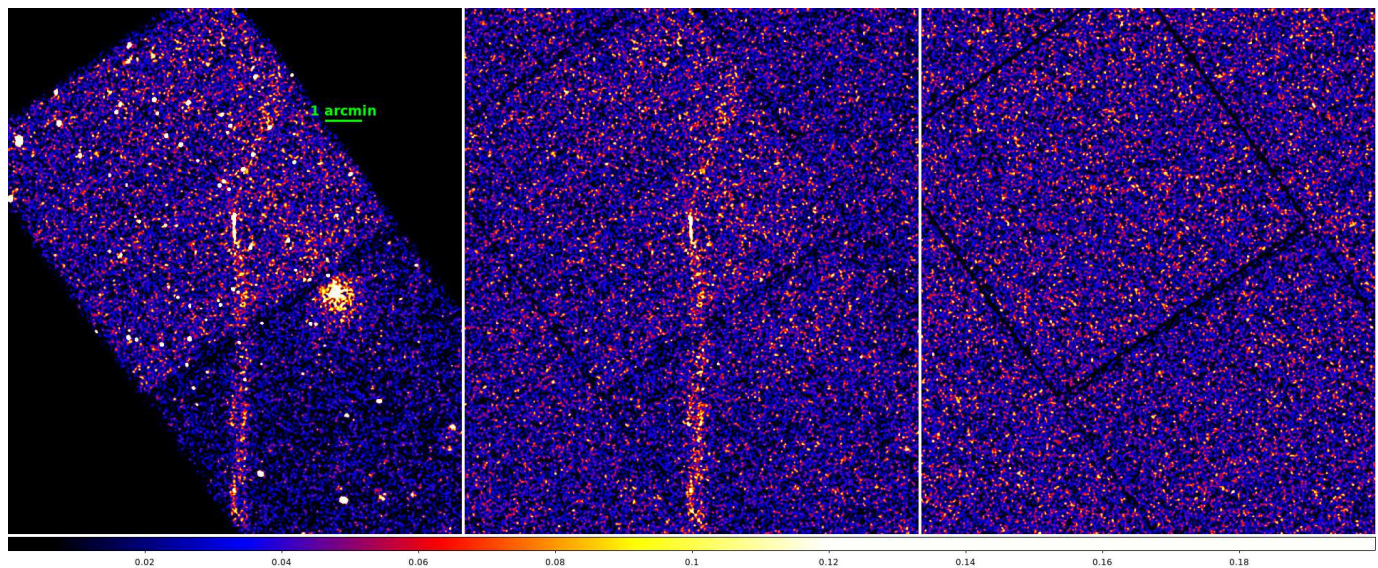}
\caption{Results of our image reduction. {\it Left panel}: Initial {\it Chandra} image. {\it Central panel}: We applied the analysis reported in
  Section \ref{flatten} to exclude field sources and flatten the {\it Chandra} background. In order to compare it with the simulated background, unexposed parts are filled
  as reported in section \ref{flatten}. {\it Right panel}: The simulated {\it Chandra} background used to evaluate the
  significance of segments found through the MRHT (see Figure \ref{hough_histo_60}).
       }
        \label{field}
\end{figure*}

As a final step, we added simulated counts to the unexposed areas following the distribution of the background of CCD 3 (figure \ref{field}).
This distribution is also used to create the simulated background image that we exploited for statistical analysis in our MRHT.

\subsection{Application of the MRHT to our {\it Chandra} data} \label{appfin}

In order to take into account the {\it Chandra} PSF, we re-binned the real and simulated images to have pixels of 1\farcs476.
We run our modified RHT on these 0.3--5 keV, exposure-corrected, binned images (real image and background-simulated image).

For each pixel $(x_i,y_i)$ within a circle centered at 20$^h$55$^m$47\fs375 +25$^{\circ}$37 48\farcs240 and of 14\arcmin\ diameter, our tool ran the HT in sub-circles of a given diameter and centered in $(x_i,y_i)$.
We ran our tool on simulated images for circles with diameters D$_W$ of 10, 20, 40, 60, 80, 100, 120, 140, 160, 180, 200, 240, 280, 320, 360, 400, 500, 600, and 700 pixels
(from 15\arcsec\ to 17\arcmin).
The simulation of the image was repeated 1000 times, and the results were averaged.
We made a quantization on the possible values of $\theta$ (from 0 to $\pi$) for each D$_W$ following the canonical \citep{cla14}

\begin{eqnarray}
  n_{\theta} = \pi \frac{\sqrt{2}}{2}(D_w-1)
.\end{eqnarray}

\begin{figure}
\centering
\includegraphics[width=9cm]{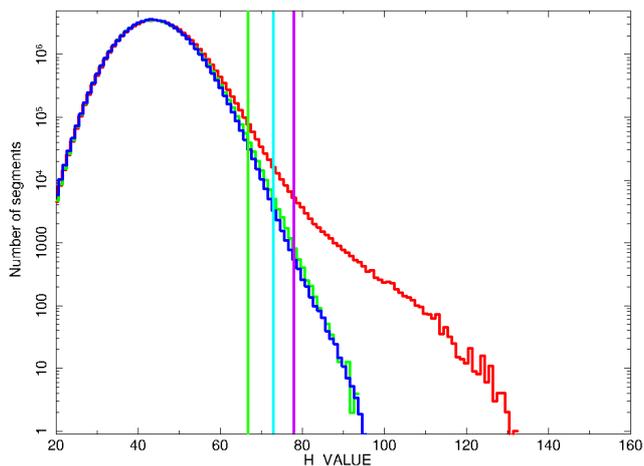}
\caption{Distribution of H-values for $D_w=40$ pixels
    ($\sim1\arcmin$). The red histogram represents the distribution
  of the real image. The green histogram represents the
    distribution of the real background, clearly overlapping with the
    blue histogram,  obtained from the simulated background, as expected.
  We highlight three different H-values with vertical lines, from left
  to right: 3$\sigma$ (green), 3.5$\sigma$ (cyan), and 4$\sigma$
  (violet), representing the chance probability of a spurious
  detection, as obtained from the simulations. The tail in the real image is apparent.
        }
        \label{hough_histo_60}
\end{figure}

We thus extracted the histogram of H-values of the simulated background for each D$_w$ (see figure \ref{hough_histo_60}).
In Figure \ref{hough_histo_60} we also report the histogram of H-values obtained for the real image background, which is consistent with the simulated one.
For each H$_{0}$-value we can evaluate in the simulation the number of spurious detections above H$_{0}$
renormalized over the entire number of detections. This represents the chance probability of a spurious detection $P(H)$ in case of a homogeneous background, which can be used
in the real image. Although the distribution is only quasi-Gaussian, as a first approximation we use the corresponding Gaussian $\sigma$ instead of the probability.

\begin{figure*}
\centering
\includegraphics[width=18cm]{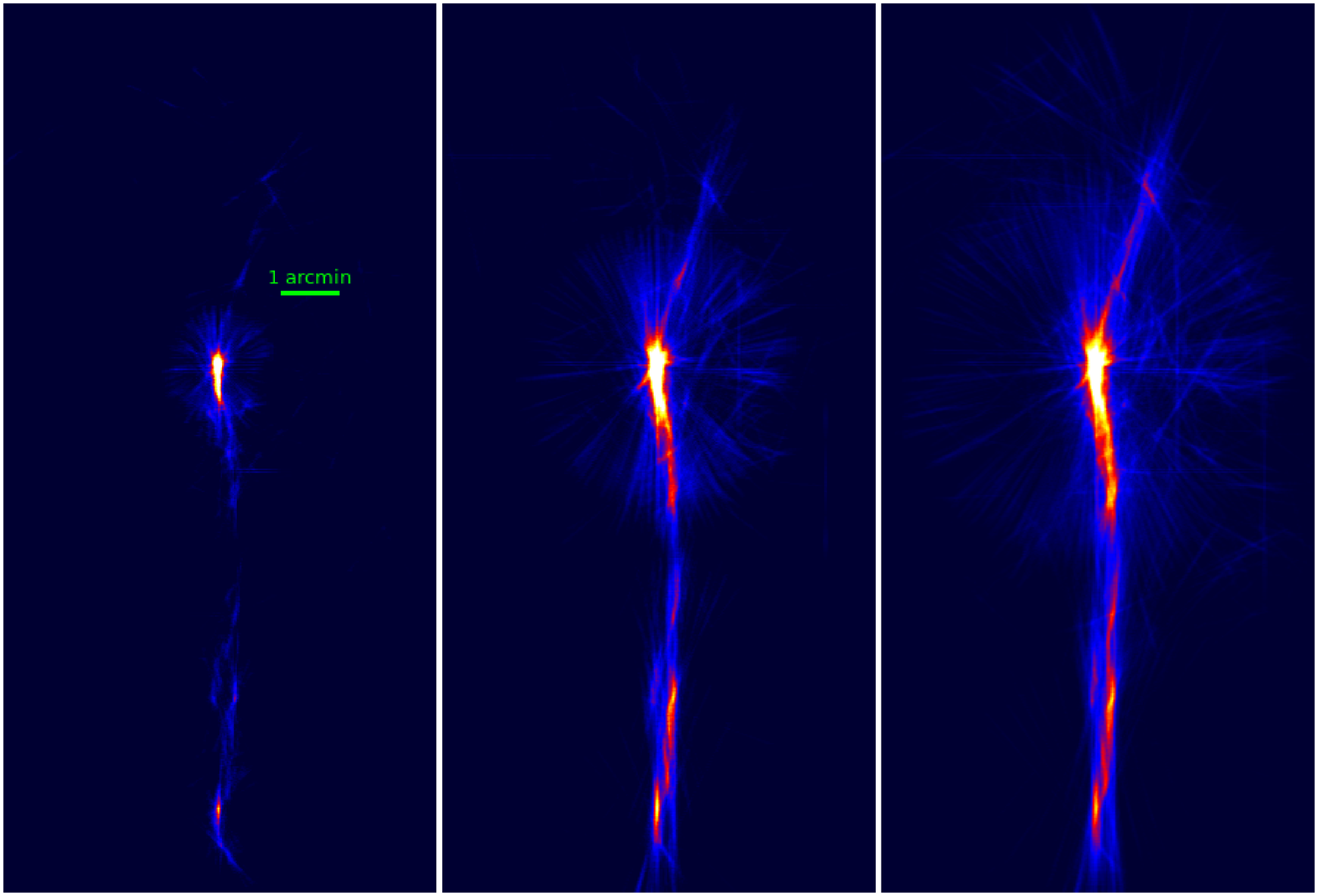}
\caption{MHRT back-projection images for different $D_w$ ($\sim1\arcmin$, $\sim2\farcm5$ and $\sim3\farcm5$). From this image, the different shapes that the nebula assumes
  using different segment lengths for the analysis are apparent. Each segment analysis best describes the nebular shape behavior at its length scale. Each box is 6\arcmin$\times$10\arcmin.
        }
        \label{hough_image}
\end{figure*}

As the final step, we ran the modified RTH on the real image, using the same prescriptions as before. We selected all
the segments with an H-value corresponding to a chance probability $>3.5\sigma$, listing $x_c,y_c,\theta,D_w,S(H),H_{val}$. This allows
for a post-selection and analysis of high-significance segments or with a given length.
We also obtain the back-projection image for each diameter and the total diameter (see figure \ref{hough_image}).\\
In order to analyze the two nebular features separately, we ran the modified RHT for the main nebula and secondary nebula separately.
Thus, for each of these runs, the region containing the excluded feature is cheesed and replaced using the expected background distribution, as in the case of point-like sources and galaxy cluster.
Finally, we also ran the script excluding both features to have the analysis of the background distribution. This revealed
some slightly preferred angles due to unresolved point sources and residuals from the exposure map and instrumental corrections: this is therefore
treated as a background for the MRHT results for the nebulae, where possible. Moreover, we only considered the segments with centers
around the position of each  of the two nebulae: a box of 150\arcsec$\times$560\arcsec\ for the main nebula and 75\arcsec$\times$280\arcsec\ for the secondary nebula.
The regions extend starting from the pulsar, following the main angles of the nebulae.

\end{appendix}

\end{document}